\documentstyle[fleqn,run2col,epsfig]{article}

\newcommand{\PO}{I\!\!P}
\newcommand{\alphapom}{\alpha_{I\!\!P}}
\newcommand{\RO}{I\!\!R}
\newcommand{\xpom}{x_{\PO}}
\newcommand{\ffig}[4]{\begin{figure}[ht]\vfill\begin{center}
\mbox{\epsfig{figure=#1,height=#2}}\caption{#3}\label{#4}
\end{center}\vfill\end{figure}}

\newcommand{\AmS}{{\protect\the\textfont2
  A\kern-.1667em\lower.5ex\hbox{M}\kern-.125emS}}

\title{QCD analysis of the diffractive structure functions measured at HERA
and factorisation breaking at Tevatron}

\author{C. Royon \address{DAPNIA/SPP, Commissariat \`a l'Energie Atomique, Saclay, 
        \\ 
        F-91191 Gif-sur-Yvette Cedex}}%
       
\begin{document}

\begin{abstract}
The 1994 data published by the H1 collaboration are
compared with models based on Regge phenomenology.
The $x_{\PO}$ dependence of
the data can be described 
in a model based on the exchange of a dominant
diffractive (pomeron) trajectory with additional sub-leading reggeon 
contributions.
The dynamics of the Pomeron structure is 
studied within the framework of perturbative QCD and new parton
distributions are obtained. These parton distributions will allow a 
direct test of factorisation breaking at Tevatron.
\end{abstract}

\maketitle

\vspace{-5mm}
\section{Regge parametrization}
The 1994 data are first investigated in the framework of a
Regge phenomenological model \cite{F2d94}.
The 1994 data are subjected to a fit in which a
single factorisable trajectory ($\PO$) is exchanged such that:
\begin{eqnarray}
F_2^{D(3)}(Q^2,\beta,x_{\PO})=
f_{\PO / p} (x_{\PO}) F_2^{\PO} (Q^2,\beta) \ .
\label{ff1}
\end{eqnarray}
In this parameterization,
$F_2^{\PO}$ can be interpreted as the structure function of the 
pomeron \cite{IS}. The value of $F_2^{\PO}$ 
is treated as a free parameter at each point
in $\beta$ and $Q^2$.
The pomeron flux takes a Regge form with a linear
trajectory $\alpha_{\PO}(t)=\alpha_{\PO}(0)+\alpha^{'}_{\PO} t$, such
that
\begin{eqnarray}
f_{\PO / p} (x_{\PO})= \int^{t_{min}}_{t_{cut}} \frac{e^{B_{\PO}t}}
{x_{\PO}^{2 \alpha_{\PO}(t) -1}} {\rm d} t \ ,
\label{flux}
\end{eqnarray}
where $|t_{min}|$ is the minimum kinematically allowed value of $|t|$ and
$t_{cut}=-1$ GeV$^2$ is the limit of the measurement. The value of
$\alpha_{\PO}(0)$ is a free parameter and 
$B_{\PO}$ and $\alpha^{'}_{\PO}$ are taken from hadron-hadron data
\cite{F2d94}. 
The fit with a single trajectory
does not give a good description of the data 
in the same way as it is observed at
$Q^2 = 0$ \cite{gammap} that secondary trajectories in addition to the
pomeron are required to describe diffractive $ep$ data.

A much better fit is obtained 
when both a 
leading ($\PO$) and a sub-leading ($\RO$) trajectory  are considered in the same
way as in formula (\ref{ff1}),
where the values of $F_2^{\PO}$ and $ F_2^{\RO}$ 
are treated as free parameters at each point
in $\beta$ and $Q^2$,
$\alpha_{\PO}(0)$ and $\alpha_{\RO}(0)$ being two free parameters.
The flux factor 
for the secondary trajectory takes the same form as equation~(\ref{flux}),
with $B_{\RO}$, and $\alpha^{'}_{\RO}$ again taken from hadron-hadron data
\cite{F2d94}. This fit yields to the following 
value of  $\alphapom(0) = 1.203 \pm 0.020 \ ({\rm stat.}) 
\pm 0.013 \ ({\rm syst.}) ^{+0.030}_{-0.035} \ ({\rm model})$ \cite{F2d94}
and is significantly larger than values extracted from
soft hadronic data ($\alpha_{\PO} \sim 1.08$).
The quality of the fit is similar
if interference between the two trajectories is introduced.

\vspace{-3mm}
\section{QCD fits and the structure of the Pomeron}
It has been suggested that
the $Q^2$ evolution of the Pomeron structure function 
may be understood in terms
of parton dynamics from perturbative QCD
where parton densities are evolved according to DGLAP \cite{dglap} equations
\cite{IS,F2d94},
using the GRV parametrization for $F_2^{\RO}$ ~\cite{GRVpion}.

For the pomeron, a quark flavour singlet distribution
($z{ {S}}_{q}(z,Q^2)=u+\bar{u}+d+\bar{d}+s+\bar{s}$)
and a gluon distribution ($z{\it {G}}(z,Q^2)$) are parameterized in terms
of coefficients $C_j^{(S)}$ and $C_j^{(G)}$ at
$Q^2_0=3$ GeV$^2$ such that~:
\begin{eqnarray}
z{\it {S}}(z,Q^2=Q_0^2)  \left[
\sum_{j=1}^n C_j^{(S)} \cdot P_j(2z-1) \right]^2
\cdot e^{\frac{a}{z-1}} \\
z{\it {G}}(z,Q^2=Q_0^2)  \left[
\sum_{j=1}^n C_j^{(G)} \cdot P_j(2z-1) \right]^2
\cdot e^{\frac{a}{z-1}}
\end{eqnarray}
where $z=x_{i/I\!\!P}$ is the fractional momentum of the pomeron carried by
the struck parton, 
$P_j(\zeta)$ is the
$j^{th}$ member in a set of Chebyshev polynomials, which are chosen such that
$P_1=1$, $P_2=\zeta$ and $P_{j+1}(\zeta)=2\zeta P_{j}(\zeta)-P_{j-1}
(\zeta)$. Some details about the fits can be found in Reference \cite{Laurent}.

A sum of $n=3$ orthonormal polynomials is used so that the input
distributions are free to adopt a large range of forms for a
given number of parameters.  The exponential factor is needed to ensure
a correct convergence close to $z$=1.

The trajectory intercepts are fixed to $\alpha_{\PO} = 1.20$ and
$\alpha_{\RO} = 0.62$. 
Only data points of H1 
with $\beta \le 0.65$, $M_X > 2$ GeV and $y \le 0.45$
are included in the fit in order to
avoid large higher twist effects and the region that may be most strongly
affected by a non zero value of $R$, the longitudinal to transverse cross-section
ratio. 

\vspace{-3mm}
\section{Results of the QCD fits}

The resulting parton densities of the Pomeron are
presented in figure~\ref{f1}. As it was noticed in the 1994 $F_2^D$ paper
\cite{F2d94}, we find two possible fits quoted here as fit 1 and fit 2. Each fit
shows a large gluonic content. The quark contribution is quite similar for both
fits, but the gluon distribution tends to be quite different at high values of $z$.
This can be easily explained as no data above $z=0.65$ are included in the fits.
Thus there is no constraint from the data at high $z$. The quark densities
is on the contrary more constrained in this region with the DGLAP evolution.
Both fits show similar
$\chi^2$ (the $\chi^2$ per degree of freedom is about 1.2)
\footnote{Fit 2 is a bit disfavoured compared to fit 1 (its $\chi^2$ by degree
of freedom is 1.3 compared to 1.2 for fit 1) and is quite instable: changing
a little the parameters modifies the gluon distribution at high $z$.}. Adding the 1995 data 
points into the fits also allows to get a better 
constraint on initial parton densities
at $Q_0^2 = 3$ GeV$^2$ compared to the fits performed with 1994 data
points alone. 
For the gluon density presented in figure ~\ref{f1}, we have determined 
that $ \frac{\delta G}{G} \simeq 25 \%$ for $z$ below 0.6.
 
The result of the fit is presented in figure~\ref{f2} together
with the experimental values
for 1994 data points~; we see on this figure the good
agreement of the QCD prediction and the data points, which supports the
validity of description of the Pomeron in terms of
partons following a QCD dynamics.

We have also tried to extend the QCD fits to lower 
$Q^2$ (below $3$ GeV$^2$) using the 1995 $F_2^D$
measurement. The $\chi^2$ of the fit turns out to increase
($\chi^2/ndf = 1.6$, adding 35 low $Q^2$ points to the 
171 points) \cite{chr}. This can be illustrated in figure 2 of 
Reference \cite{chr}
where changes of slopes
of scaling violations
for $Q^2$ below and above $3$ GeV$^2$ can be seen. It may indicate that breaking
of perturbative QCD has already occured in this region.

The idea would then  to use these parton distributions and to compare with 
the measurements at Tevatron in order to study factorisation
breaking. The roman pots which will be available in the D0
experiment at Run II will allow a direct comparison with the results obtained 
from the HERA parton distributions. It will be possible to know where
factorisation breaking takes place at Tevatron, e.g. is it at low or
high $\beta$?

\ffig{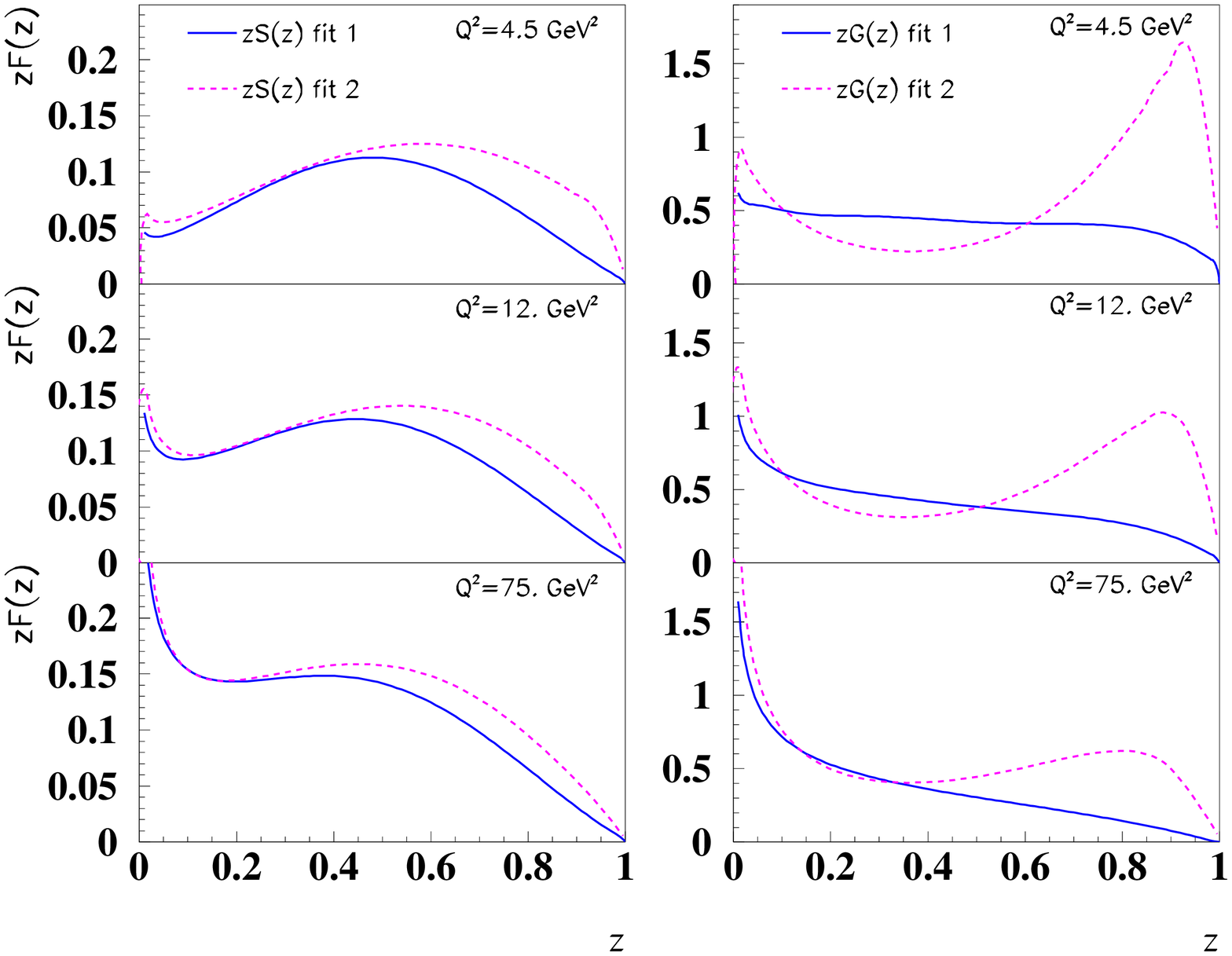}{80 mm}{
Quark flavour singlet ($zS$, left) and gluon ($zG$, right) distributions
of the pomeron deduced as a function of $z$, the fractional momentum of the
pomeron carried by the struck parton, from the fit
on 1994 data points with $Q^2 \ge 4$ GeV$^2$. Two possible fits labelled
as fit 1 and fit 2 are found
($\chi^2/ndf = 1.2$ for fit 1,and $\chi^2/ndf = 1.3$ for fit 2 with 
statistical errors only).}{f1}

\ffig{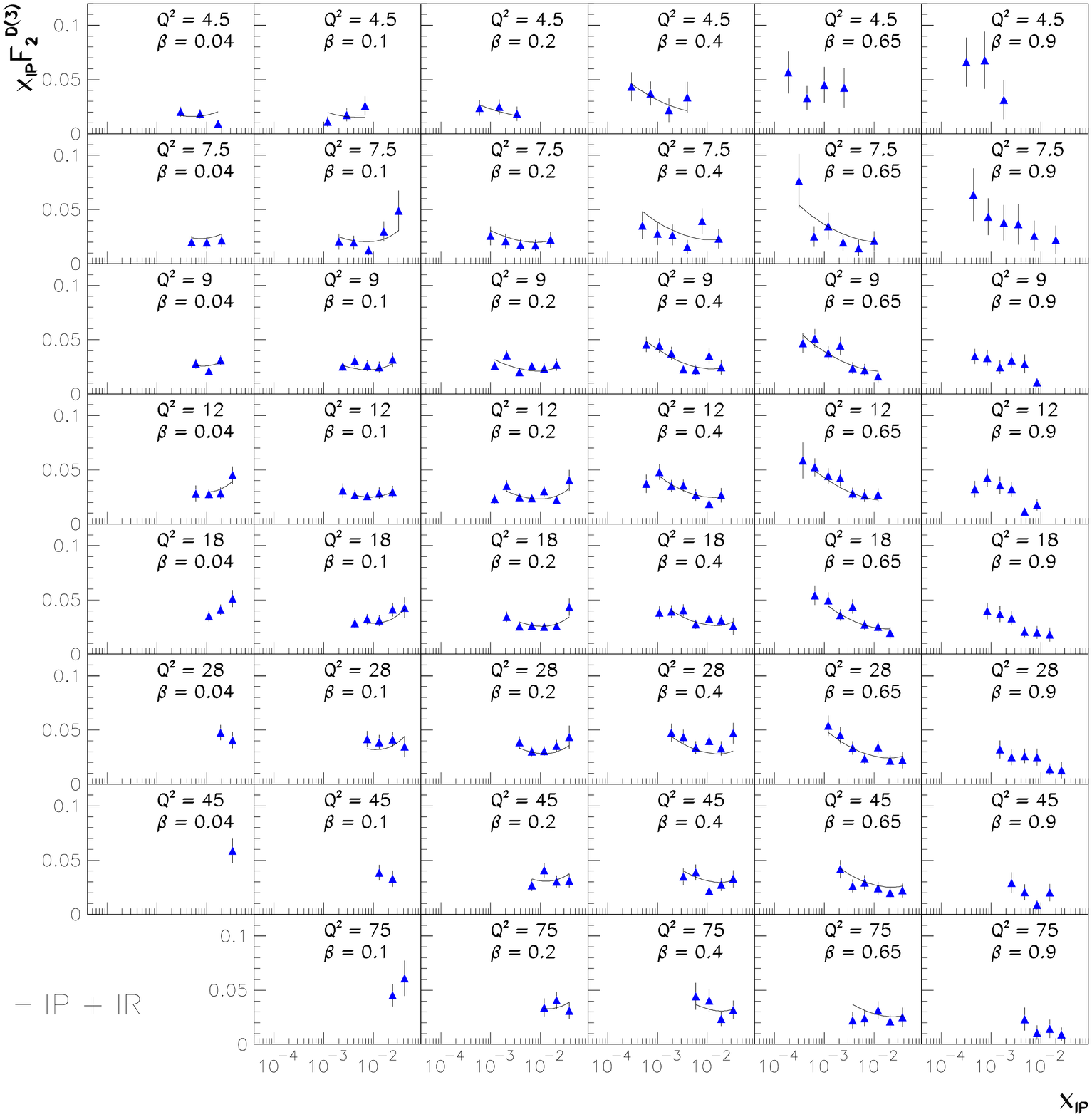}{160 mm}{
The H1 data points on $\xpom F_2^{D(3)}$ (1994)
are shown with the result
of the QCD fit described in the text; the result of the
fit is drawn only in bins
included in the minimization procedure.}{f2}

\vspace{-3mm}
\section{Acknowledgments}
The results described in the present contribution come from a fruitful
collaboration with J.Bartels, H.Jung R.Peschanski and L.Schoeffel.

\end{document}